\documentclass[10pt, conference, compsocconf]{IEEEtran}
\usepackage{algorithm,algorithmic,amsbsy,amsmath,amssymb,epsfig,bbm,mathrsfs,multirow,amsthm}
\usepackage{subfigure}

\newtheorem{theorem}{\textbf{\textsc{Theorem}}}

\begin{document}
\title{The Tradeoff Analysis in RF-Powered Backscatter Cognitive Radio Networks}
\author{Dinh Thai Hoang$^1$, Dusit Niyato$^1$, Ping Wang$^1$, Dong In Kim$^2$, and Zhu Han$^3$\\
$^1$ School of Computer Engineering, Nanyang Technological University (NTU), Singapore	\\
$^2$ School of Information and Communication Engineering, Sungkyunkwan University (SKKU), Korea		\\
$^3$ Department of Electrical and Computer Engineering, University of Houston, USA 	\vspace{-5mm}	}

\maketitle
\begin{abstract}
In this paper, we introduce a new model for RF-powered cognitive radio networks with the aim to improve the performance for secondary systems. In our proposed model, when the primary channel is busy, the secondary transmitter is able either to backscatter the primary signals to transmit data to the secondary receiver or to harvest RF energy from the channel. The harvested energy then will be used to transmit data to the receiver when the channel becomes idle. We first analyze the tradeoff between backscatter communication and harvest-then-transmit protocol in the network. To maximize the overall transmission rate of the secondary network, we formulate an optimization problem to find time ratio between taking backscatter and harvest-then-transmit modes. Through numerical results, we show that under the proposed model can achieve the overall transmission rate higher than using either the backscatter communication or the harvest-then-transmit protocol. 
\end{abstract}

{\it Keywords-} Cognitive radio networks, ambient backscattering, RF energy harvesting, convex optimization. 
 
\section{Introduction}
Recently, RF energy harvesting technique has been integrated and implemented in cognitive radio networks (CRNs). This leads to a new type of networks, called RF-powered CRNs. In these networks, secondary users can harvest RF energy when a primary channel is busy, and use the energy to transmit data when the primary channel is idle~\cite{Park2013Optimal, Hoang2014Opportunistic}. This is referred to as the harvest-then-transmit protocol/mode. There are many advantages of RF energy harvesting in CRNs as discussed in~\cite{Hoang2015Performance}. However, for RF-powered CRNs, when the channel idle probability is low, i.e., the channel is mostly occupied by primary users, the secondary transmitters have less opportunity to transmit data, resulting in a low overall transmission rate for secondary networks. Therefore, there is a need to overcome this shortcoming. 

Ambient backscatter communication~\cite{LiuAmbient2013, Kellogg2014Wi-fi} has been introduced as a new communication method. The technique allows wireless data transmission between two wireless nodes using ambient signals without needing a standard form of energy supply and storage. In ambient backscatter communication, when a transmitter wants to communicate with a receiver, the transmitter will backscatter signals received from a signal source, e.g., a TV tower, to its receiver. The receiver then can decode and obtain useful information from the transmitter. However, similar to traditional RF-powered CRNs, the performance of backscatter communication greatly depends on the ambient signals. Specifically, when the idle channel probability is high, the performance of the ambient backscatter communication is low due to limited time to backscatter. Therefore, in this paper, we propose a novel model which utilizes the advantages of both backscatter communication and harvest-then-transmit protocol in RF-powered CRNs.

In particular, we consider an RF-powered CRN with the backscatter communication capability. In the network, the secondary transmitter (ST) is able not only to harvest energy from radio signals, but also to backscatter these signals to its receiver for data transmission. As highlighted in~\cite{Zhang2014Enabling}, backscatter communication and energy harvesting cannot practically be performed at the same time. If the ST performs backscatter communication, the RF carrier wave is being modulated which can significantly reduce the amount of harvested energy, and mostly it is not sufficient to transmit data. Clearly, when the channel is mostly busy, the ST should use backscatter mode to transmit data. By contrast, when the channel is less busy, the ST should use the harvest-then-transmit mode. This leads to an important question of how to tradeoff the time for using backscatter and harvest-then-transmit modes such that the overall transmission rate of the secondary user is maximized. Here, the overall transmission rate refers to the total rate from both backscatter and harvest-then-transmit modes. 
 
\begin{figure*}[t]
\centering
\includegraphics[scale=0.62]{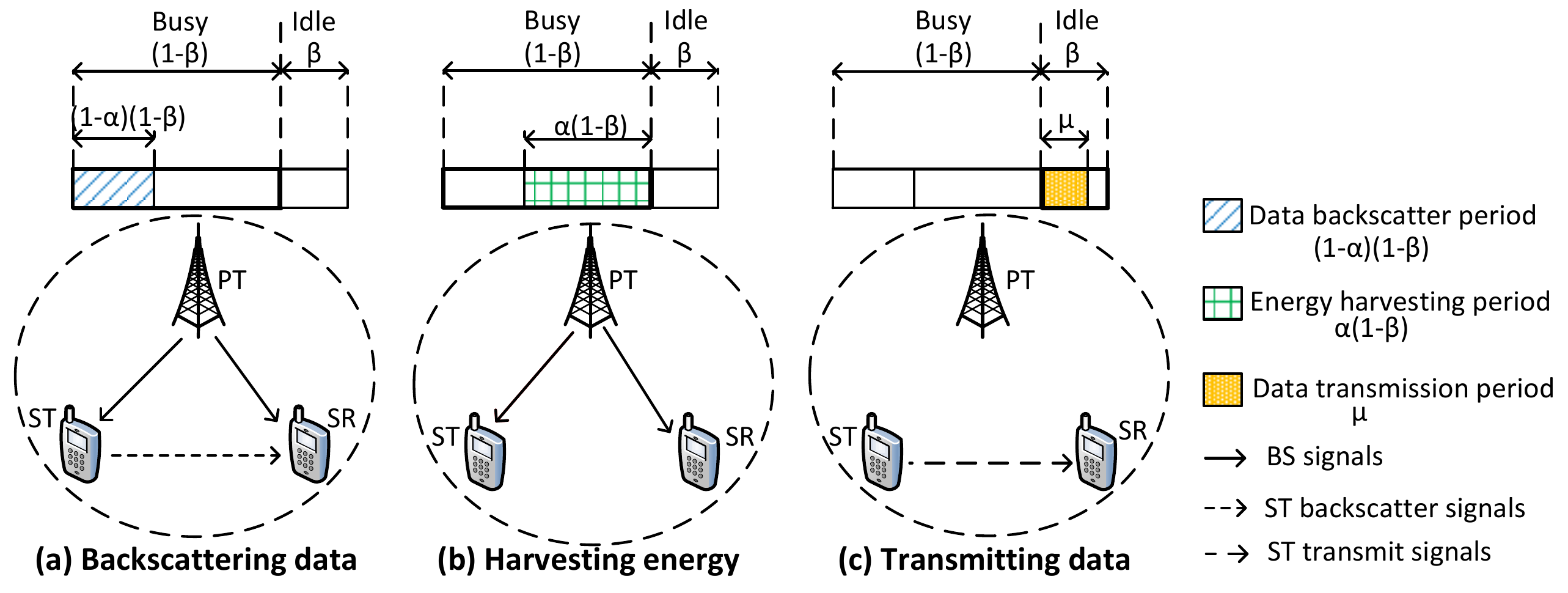}
\caption{RF-powered cognitive radio network with ambient backscatter communication.}
\label{System_Model}
\end{figure*}

The time tradeoff problem for wireless powered communication networks was studied in few work in the literature. For example, in~\cite{Ju2014Throughput}, the authors studied the tradeoff between the wireless energy transfer and wireless information transmission for wireless powered communication networks by introducing the harvest-then-transmit protocol with the aim to maximize the transmission rate for the network. In~\cite{Rakovic2015Optimal}, an optimal tradeoff time between the energy harvesting phase and the data transmission phase for an underlay CRN was investigated through adopting the convex optimization technique. Extending from~\cite{Rakovic2015Optimal}, the authors in~\cite{Yin2014Optimal} considered a cooperation scenario for primary users (PUs) and secondary users (SUs). The SUs need to determine not only how much time for energy harvesting, but also how much power for PUs' data relay or data transmission to allocate.

In this paper, we analyze the tradeoff between energy harvesting and backscatter communication for an overlay RF-powered CRN. The main aim is to improve the overall transmission rate for the secondary network. We formulate an optimization problem to obtain the optimal time to perform energy harvesting and backscatter communication when the channel is busy. We show that the problem is convex, and hence any tool from convex optimization can be used to obtain a globally optimal solution. Through numerical results, we demonstrate that our proposed solution can significantly improve the performance for the secondary network compared with baseline methods. To the best of our knowledge, this is the first work that proposes the idea of integrating the ambient backscatter communication with wireless powered CRNs. Moreover, we are the first that introduce the tradeoff analysis in RF-powered backscatter CRNs.

\section{System Model}

\subsection{Network Setting}

We study an RF-powered backscatter CRN composed of a primary transmitter (PT), and a secondary transmitter (ST) communicating with a secondary receiver (SR). The ST is equipped with an RF energy harvesting module and a backscatter circuit in order to harvest RF energy and backscatter radio signals, respectively. The ST can also transmit data as normal wireless transmission. When the PT, e.g., an amplitude modulated (AM) broadcasting base station (BS) or a TV tower, transmits RF signal to its primary receiver (PR), the primary channel is busy. At the same time, the ST can either harvest energy and store it in the energy storage or backscatter the signal for data transmission~\cite{LiuAmbient2013}. The harvested energy is used for direct wireless data transmission to the SR when the primary channel is idle. This is referred to as the harvest-then-transmit mode while the other is referred to as the backscatter mode. We assume that the SR perfectly knows the mode of the ST and applies corresponding demodulators to extract useful information.

\subsection{Tradeoff in RF-Powered Backscatter Cognitive Radio Network}

\begin{figure}[!]
\centering
\epsfxsize=2.6 in \epsffile{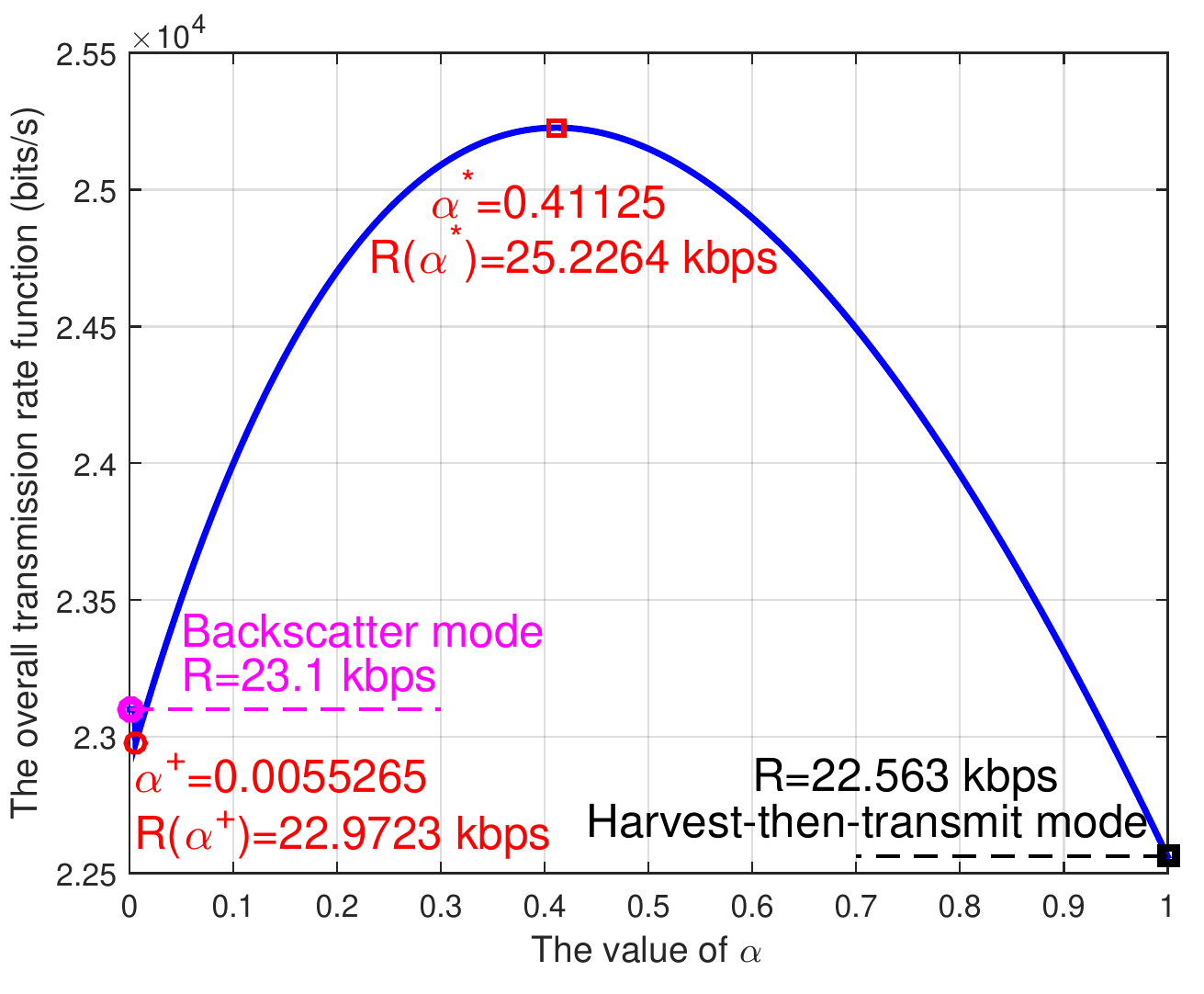}
\caption{The optimal value of $\alpha$.}
\label{fig:Motivate}
\end{figure}

In the proposed system, when the PT transmits signals, i.e., the primary channel is busy, the ST can transmit data to the SR using backscatter communication (Fig.~\ref{System_Model} (a)) or harvest energy (Fig.~\ref{System_Model} (b)). Let $\beta$ denote the normalized channel idle period and $(1-\beta)$ denote the normalized channel busy period (as shown in Fig.~\ref{System_Model}). When the channel is busy, $\alpha$ denotes the time ratio for energy harvesting, and $(1-\alpha)$ denotes the time ratio for backscatter communication. The energy harvested during the time ratio $\alpha$ will be used for direct data transmission during the idle channel period. We observe that there is a tradeoff between the time ratio for backscatter communication and energy harvesting. As shown in Fig.~\ref{fig:Motivate}~\footnote{The parameter setting for obtaining the result in Fig.~\ref{fig:Motivate} is provided in Section~\ref{sec:Performance Evaluation}.}, when $\alpha$ is small, i.e., the ST spends much time for backscatter communication, the overall transmission rate is small. This is from the fact that the ST cannot fully utilize the channel idle period for direct data transmission due to small amount of energy harvested. As $\alpha$ increases, the overall transmission rate increases since more harvested energy can be used to transmit more data. However, when the ST spends much time for energy harvesting, i.e., $\alpha$ is high, the overall transmission rate decreases since the channel idle period is limited while the backscatter communication is not efficiently used during the channel busy period.

Clearly, the ST can achieve the optimal overall transmission rate by balancing between the backscatter communication and energy harvesting during the busy channel period. In particular, there is an optimal value for $\alpha^*$ which we aim to achieve by formulating and solving an optimization problem presented in the following sections.


\section{Problem Formulation and Proposed Solution} 

\subsection{Problem Formulation} 
\label{sec:Problem Formulation}

We aim at maximizing the overall transmission rate of the secondary network which is the number of information bits transmitted by the ST per time unit. We denote $R$ as the overall transmission rate which is obtained as follows:
\begin{equation}
	R = R_{\mathrm{b}} + R_{\mathrm{h}} 	,
\end{equation}
where $R_{\mathrm{b}}$ and $R_{\mathrm{h}}$ are the numbers of transmitted bits using the backscatter mode and the harvest-then-transmit mode in a time unit, respectively. 

\subsubsection{Backscatter mode} 
\label{subsec:Backscatter mode}

In the following, we explain how SR can receive information through using ambient backscatter and how to control backscatter transmission rate between ST and SR. 

\paragraph{Extracting backscatter information from ambient signals} 

We briefly describe the method used by the SR to extract information from the ST through the ambient backscatter communication. For more details, the readers are referred to~\cite{LiuAmbient2013}. The core idea of backscatter communication is that the ST backscatters information at a lower rate than that of ambient signals, e.g., signals from the PT. Thus, the SR is able to distinguish such two signals by using the averaging mechanisms. In particular, the authors in~\cite{LiuAmbient2013} presented a simple circuit diagram to demodulate the information for the SR. There are two stages, i.e., averaging stage and compute threshold stage. In the first stage, the SR smooths and averages the natural variations in the PT signals. The output of the averaging stage yields two signal levels, corresponding to the voltage $V_0$ (bit `0') and the voltage $V_1$ (bit `1') for $V_1 > V_0$. Then, in the second stage, the SR computes the threshold between these two levels, which is the average of the two signal levels, i.e., $\frac{V_0 + V_1}{2}$. If the received signal is greater than the threshold, the SR concludes that the received signal is $V_1$, and $V_0$ otherwise. Finally, the comparator takes two voltages as inputs and generates a bit `0' or `1' accordingly. 


\paragraph{Transmission rate of backscatter mode} 

It is shown in~\cite{LiuAmbient2013} that the transmission rate of the ambient backscatter communication depends on the setting of the RC circuit elements. For example, to transmit data at the transmission rate of 1kbps and 10kbps, the values of circuit elements, i.e., $R_1$, $R_2$, $C_1$, and $C_2$, are set at (150 $k\Omega$, 10 $M\Omega$, 4.7 $nF$, 10 $nF$) and (150 $k\Omega$, 10 $M\Omega$, 680 $pF$, 1 $\mu F$), respectively. Therefore, let $B_{\mathrm{b}}$ denote the transmission rate of the ambient backscatter communication, the total number of bits transmitted using the backscatter mode for the RF-powered backscatter CRN is expressed as follows: 
\begin{equation}
\label{eq:R_b}
	R_{\mathrm{b}} = (1-\beta) (1-\alpha)	B_{\mathrm{b}}.
\end{equation}
Here, we note that through real implementations in~\cite{LiuAmbient2013}, when the ST backscatters signals to the SR, the ST still can harvest energy from RF signals. Although the amount of harvested energy is not enough to transmit data (when the channel is idle), it is sufficient to sustain backscatter operations of the ST. Therefore, in~(\ref{eq:R_b}), there is no need to consider the circuit energy consumption for the backscatter mode.

\subsubsection{Harvest-then-transmit mode} 
\label{subsec:Harvest-then-transmit mode}

This mode includes two phases. First, the ST harvests energy from the PT in the energy harvesting period. Then, the ST will use the harvested energy to transmit data in the data transmission period. In the following, we show the amount of energy that the ST can harvest in the first phase and the number of bits transmitted in the second phase.

\paragraph{Harvesting energy} 

From Friis equation~\cite{Balanis2012}, we can determine the harvested RF power from the PT for the ST in a free space as follows:
\begin{equation}
\label{eq:Friis}
	P_{\mathrm{R}} = \delta P_{\mathrm{T}} \frac{G_{\mathrm{T}} G_{\mathrm{R}} \lambda^2}{(4 \pi d)^2}	,
\end{equation}
where $P_{\mathrm{R}}$ is the ST's harvested power, $P_{\mathrm{T}}$ is the PT transmission power, $\delta$ is the energy harvesting efficiency, $G_{\mathrm{T}}$ is the PT antenna gain, $G_{\mathrm{R}}$ is the ST antenna gain, $\lambda$ is the emitted wavelength, and $d$ is the distance between the PT and the ST. We then derive the total amount of harvested energy over the energy harvesting period $\alpha (1-\beta)$ as follows:
\begin{equation}
	E_{\mathrm{h}} =  \alpha (1-\beta) P_{\mathrm{R}} = \alpha (1-\beta) \delta P_{\mathrm{T}} \frac{G_{\mathrm{T}} G_{\mathrm{R}} \lambda^2}{(4 \pi d)^2}.
\end{equation}

\paragraph{Transmitting data} 

After harvesting energy in the first phase, the ST will use all harvested energy subtracted by the circuit energy consumption to transmit data over the data transmission period $\mu$ when the channel is idle. Let $P^{\mathrm{tr}}$ denote the transmission power of the ST in the data transmission period $\mu$ ($\mu \in [0,\beta]$ as shown in Fig.~\ref{System_Model} (c)) when the channel is idle. Thus, $P^{\mathrm{tr}}$ can be obtained from
\begin{equation}
\label{eq:P^{tr}}
	P^{\mathrm{tr}} = \frac{E_{\mathrm{h}} - E_{\mathrm{c}}}{\mu}  ,
\end{equation}
where $E_{\mathrm{h}}$ is the total harvested energy and $E_{\mathrm{c}}$ is the circuit energy consumption. From~\cite{Huang2012Decentralized}, given the transmit power $P^{\mathrm{tr}}$, the transmit data rate can be determined as follows:
\begin{equation}
r_h = \kappa W  \log_2 \left(	1 + \frac{P^{\mathrm{tr}}}{P_0}		\right),
\end{equation}
where $\kappa \in [0,1]$ is the transmission efficiency, $W$ is the bandwidth of the primary channel, and $P_0$ is the ratio between the noise power $N_0$ and the channel gain coefficient $h$, i.e., $P_0 = \frac{N_0}{h}$.

Then, the number of transmitted bits per time unit using the harvest-then-transmit mode is given by
\begin{equation}
\label{eq:trans_data_rate}
	R_{\mathrm{h}} = \mu \kappa W \log_2 \left(	1 + \frac{P^{\mathrm{tr}}}{P_0}		\right)	.
\end{equation}

Here, since $R_{\mathrm{h}}$ in~(\ref{eq:trans_data_rate}) must be non-negative, $P^{\mathrm{tr}}$ in~(\ref{eq:P^{tr}}) must be also non-negative. Consequently, from~(\ref{eq:P^{tr}}), we have the following condition:
\begin{eqnarray}
\label{eq:condition_of_R_h}
	E_{\mathrm{h}}	& = 	&  \alpha (1-\beta) P_{\mathrm{R}} \geq E_{\mathrm{c}}, \phantom{5} \text{it means}	\\
	\alpha 			& \geq	&	\frac{E_{\mathrm{c}}}{(1-\beta) P_{\mathrm{R}}}	.
\end{eqnarray}
We denote $\alpha^{\dagger}=\frac{E_{\mathrm{c}}}{(1-\beta) P_{\mathrm{R}}}$ as the minimum energy harvesting time to obtain enough energy for supplying the circuit of the ST to use the harvest-then-transmit mode. Then, we have $\alpha \geq \alpha^{\dagger}$. Note that we have $\alpha \leq 1$. Therefore, if $\alpha^{\dagger} \leq 1$, then $R_{\mathrm{h}}$ can be greater than zero. We denote $m=\frac{(1-\beta)}{P_0 \mu} P_{\mathrm{R}}$ and $n=1-\frac{E_{\mathrm{c}}}{P_0 \mu}$, then from~(\ref{eq:trans_data_rate}) we have
\begin{equation}
R_{\mathrm{h}} =	
 	\left\{	\begin{array}{ll}
 	\mu \kappa W \log_2 (n +  m \alpha),	& \text{if} \phantom{5} \alpha^{\dagger} \leq 1 \phantom{5}\text{and}\phantom{5} \alpha^{\dagger} \leq \alpha	,	\\
	0, 										& \text{otherwise} .
	\end{array}	\right.
\label{eq:R_h}
\end{equation}
Here, we note that $m>0$ and $(n+m\alpha) >0, \forall \alpha \in [\alpha^{\dagger},1]$.

Then, the optimization problem can be formulated as in~(\ref{eq:opt_trans_bits_two_variables}) (on the top of next page).
\begin{figure*}[!]
\normalsize
\begin{equation}
\begin{aligned}
&	\max_{\alpha, \mu} R(\alpha, \mu) 	 =	
& 	\left\{	\begin{array}{ll}
	(1-\beta) (1-\alpha)	B_{\mathrm{b}} + \mu \kappa W \log_2 (n +  m \alpha),		&	\text{if} \phantom{5} \alpha^{\dagger} \leq 1 \phantom{5}\text{and}\phantom{5} \alpha^{\dagger} \leq \alpha	,	\\
	(1-\beta) (1-\alpha)	B_{\mathrm{b}}, 											&	\text{otherwise} .
	\end{array}	\right.
\end{aligned}
\label{eq:opt_trans_bits_two_variables}
\end{equation}
\hrulefill
\vspace*{-4pt}
\end{figure*}

\subsection{Proposed Solution} 

First, from~(\ref{eq:opt_trans_bits_two_variables}), when $R(\alpha, \mu)=(1-\beta) (1-\alpha) B_{\mathrm{b}}$, it is easy to show that
\begin{equation}
\max_{\alpha,\mu} R (\alpha,\mu) = R(\alpha=0) = (1-\beta)B_{\mathrm{b}}, \forall \alpha \in [0,1].
\end{equation}
Second, through Theorem~\ref{theo:opt_data_trans_period}, we will prove that when $\alpha^{\dagger} \leq 1$ and $\alpha^{\dagger} \leq \alpha$, the optimal overall transmission rate is achieved when the ST transmits data over the entire channel idle period, i.e., $\max_{\alpha,\mu} R (\alpha,\mu) = R(\alpha,\beta)$.

\begin{theorem}
\label{theo:opt_data_trans_period}
When $\alpha^{\dagger} \leq 1$ and $\alpha^{\dagger} \leq \alpha$, if we consider $R_{\mathrm{h}}$ from~(\ref{eq:R_h}) as a function of $\mu$, then $R_{\mathrm{h}}$ reaches the highest value if and only if $\mu = \beta$. In other words,
\begin{equation}
\max_{\mu} R_{\mathrm{h}} (\mu) = R_{\mathrm{h}}(\beta), \forall \mu \in [0,\beta].
\end{equation}
\end{theorem}
The proof of Theorem~\ref{theo:opt_data_trans_period} is ignored due to the limited pages. 

From Theorem~\ref{theo:opt_data_trans_period}, the optimization problem in~(\ref{eq:opt_trans_bits_two_variables}) can be rewritten with only one variable $\alpha$ as in~(\ref{eq:opt_trans_bits_one_variables}) (on the top of next page).
\begin{figure*}[!]
\normalsize
\begin{equation}
	\max_{\alpha} R(\alpha) 	 =	
	\left\{	\begin{array}{ll}
	(1-\beta) (1-\alpha)	B_{\mathrm{b}} + \beta \kappa W \log_2 (n +  m \alpha),		&	\text{if} \phantom{5} \alpha^{\dagger} \leq 1 \phantom{5}\text{and}\phantom{5} \alpha^{\dagger} \leq \alpha	\\
	(1-\beta)B_{\mathrm{b}}, 															&	\text{otherwise} .
	\end{array}	\right.
\label{eq:opt_trans_bits_one_variables}
\end{equation}
\hrulefill
\vspace*{-4pt}
\end{figure*}
After that, we give the following theorem.
\begin{theorem}
\label{theo:opt_bit_rate_function}
When $\alpha \in [\alpha^{\dagger},1]$ and $\alpha^{\dagger} \leq 1$ and the value backscatter transmission rate $B_{\mathrm{b}} \in \Big(\frac{\beta \kappa W m}{(m+n)(1-\beta)\ln 2},\frac{\beta \kappa W m}{(m \alpha^{\dagger}+n)(1-\beta)\ln 2}  \Big)$, there exists a globally optimal solution of $\alpha^* \in [\alpha^{\dagger},1]$ which maximizes $R$. 
\end{theorem}
The proof of Theorem~\ref{theo:opt_bit_rate_function} is similar to the proof of Theorem~\ref{theo:opt_data_trans_period} in Appendix~\ref{app:theo:opt_data_trans_period}. We ignore here due to the limited pages.

\begin{theorem}
\label{theo:mode_selection}
For $\alpha \in [\alpha^{\dagger},1]$ and $\alpha^{\dagger} \leq 1$, if $B_{\mathrm{b}} \geq \frac{\beta \kappa W m}{(m \alpha^{\dagger}+n)(1-\beta)\ln 2}$, then $\alpha^{*} = \alpha^{\dagger}$. Moreover, when $B_{\mathrm{b}} \leq \frac{\beta \kappa W m}{(m+n)(1-\beta)\ln 2}$, then $\alpha^{*} = 1$. 
\end{theorem}
The proof of Theorem~\ref{theo:mode_selection} is similar to the proof of Theorem~\ref{theo:opt_data_trans_period} in Appendix~\ref{app:theo:opt_data_trans_period}. We ignore here due to the limited pages.  

From Theorem~\ref{theo:opt_bit_rate_function} and Theorem~\ref{theo:mode_selection}, we show graphically the optimal solution $\alpha^* \in [\alpha^{\dagger},1]$ under the variation of $B_{\mathrm{b}}$ in Fig.~\ref{Optimal_Decisions}. Note that the convexity of the objective function $R$ is proved in Appendix~\ref{app:theo:opt_bit_rate_function} and validated in Fig.~\ref{fig:Motivate}.
\begin{figure}[htb]
\centering
\epsfxsize=3.2 in \epsffile{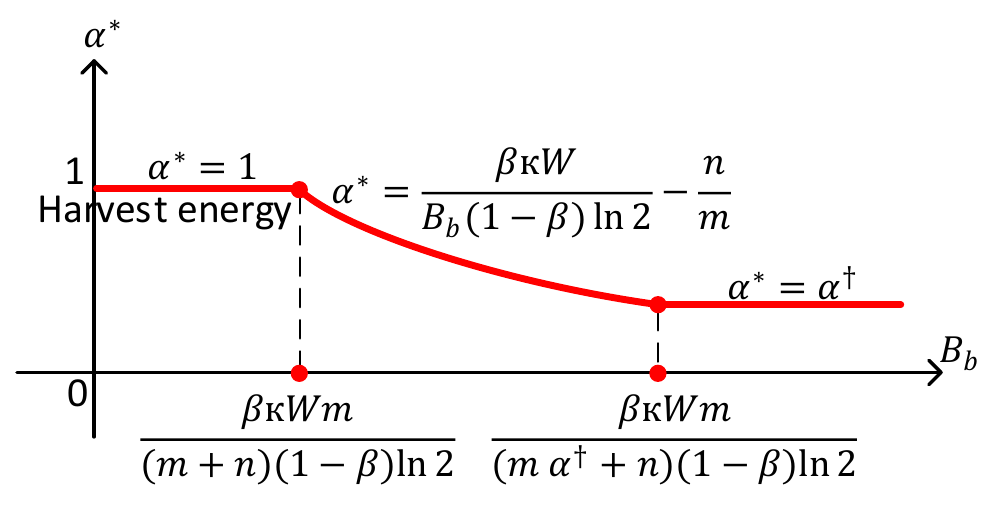}
\caption{Optimal value of $\alpha$ under the variation of $B_{\mathrm{b}}$ when $R_{\mathrm{h}} \geq 0$.}
\label{Optimal_Decisions}
\end{figure}

Finally, we can derive the maximum value of $R$ as in~(\ref{Optimal_R}) (on the top of next page).
\begin{figure*}[!]
\normalsize
\begin{equation}
R_{\max} =
 	\left\{	\begin{array}{ll}
	\max\big[(1-\beta)B_{\mathrm{b}}, (1-\beta) (1-\alpha^{*})	B_{\mathrm{b}} + \beta \kappa W \log_2 (n +  m \alpha^{*})\big], 		&	\text{if} \phantom{5} \alpha^{\dagger} \leq 1 \phantom{5}\text{and}\phantom{5} \alpha^{\dagger} \leq \alpha \\
	(1-\beta)B_{\mathrm{b}}, 						&	\text{otherwise}.
	\end{array}	\right.
\label{Optimal_R}
\end{equation}
\hrulefill
\vspace*{-4pt}
\end{figure*}

\section{Performance Evaluation} 
\label{sec:Performance Evaluation}

\subsection{Experiment Setup} 
In the RF-powered backscatter CRN under our consideration, the PT is an FM radio station. The bandwidth and the frequency of the FM signals are set at $100$kHz and $100$MHz, respectively. The idle channel probability is $0.3$. Unless otherwise stated, the transmission power of the PT that broadcasts BM signals is set at $10$kW, and the backscatter transmission rate are set at $33$kbps. The PT antenna gain and ST antenna gain are set at 6dbi as in~\cite{Kim2010Reverse}, and the circuit power consumption is set at -35dbm. Similar to~\cite{LiuAmbient2013}, the distance from PT to ST is assumed to be around 6.7 miles while the distance between ST and SR is within 1 meter. The energy harvesting efficiency and data transmission efficiency are set at $0.6$. 

\subsection{Numerical Results} 

\begin{figure}[!]
\begin{center}
$\begin{array}{c} 
\epsfxsize=2.6 in \epsffile{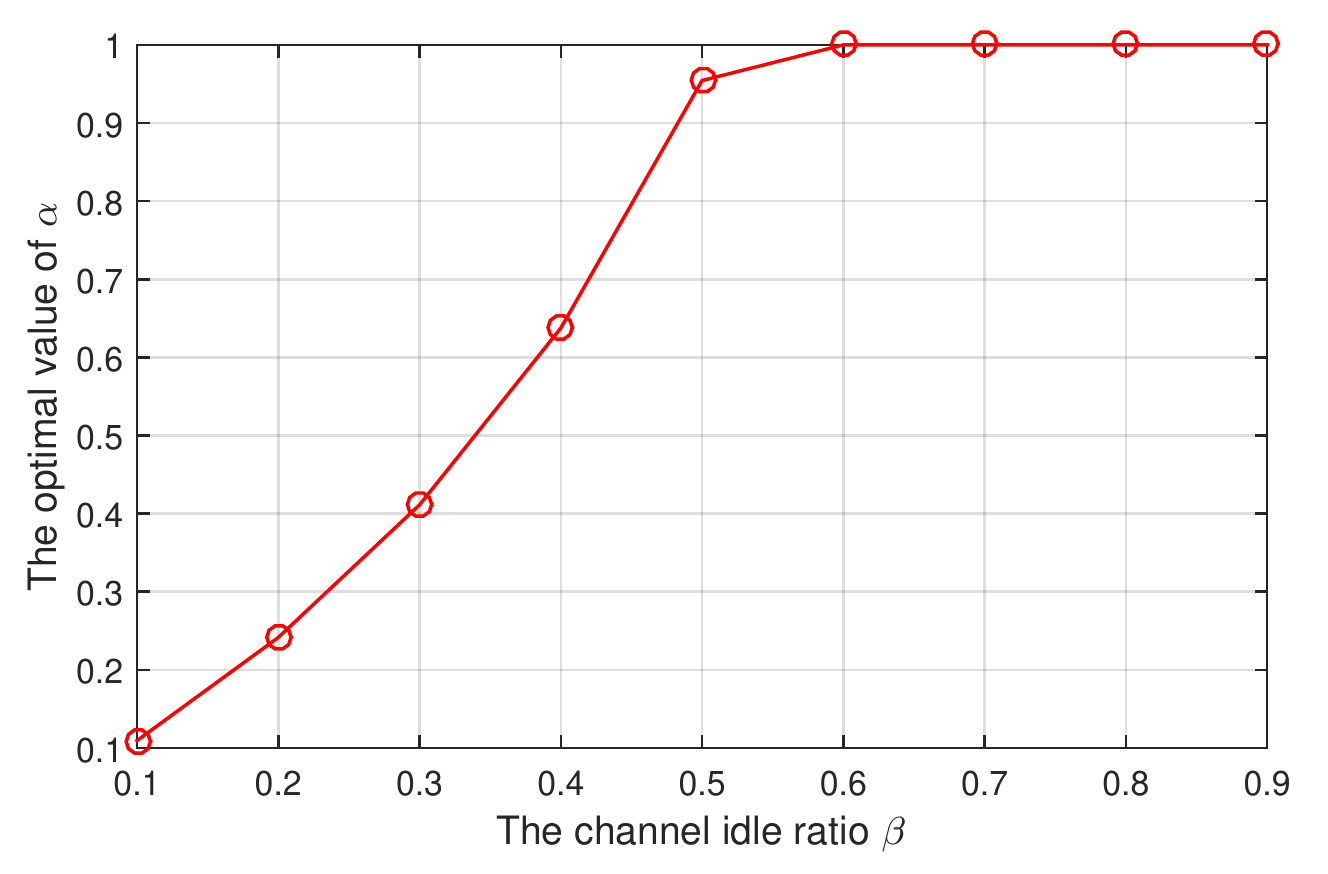}   \\
(a) \\
\epsfxsize=2.6 in \epsffile{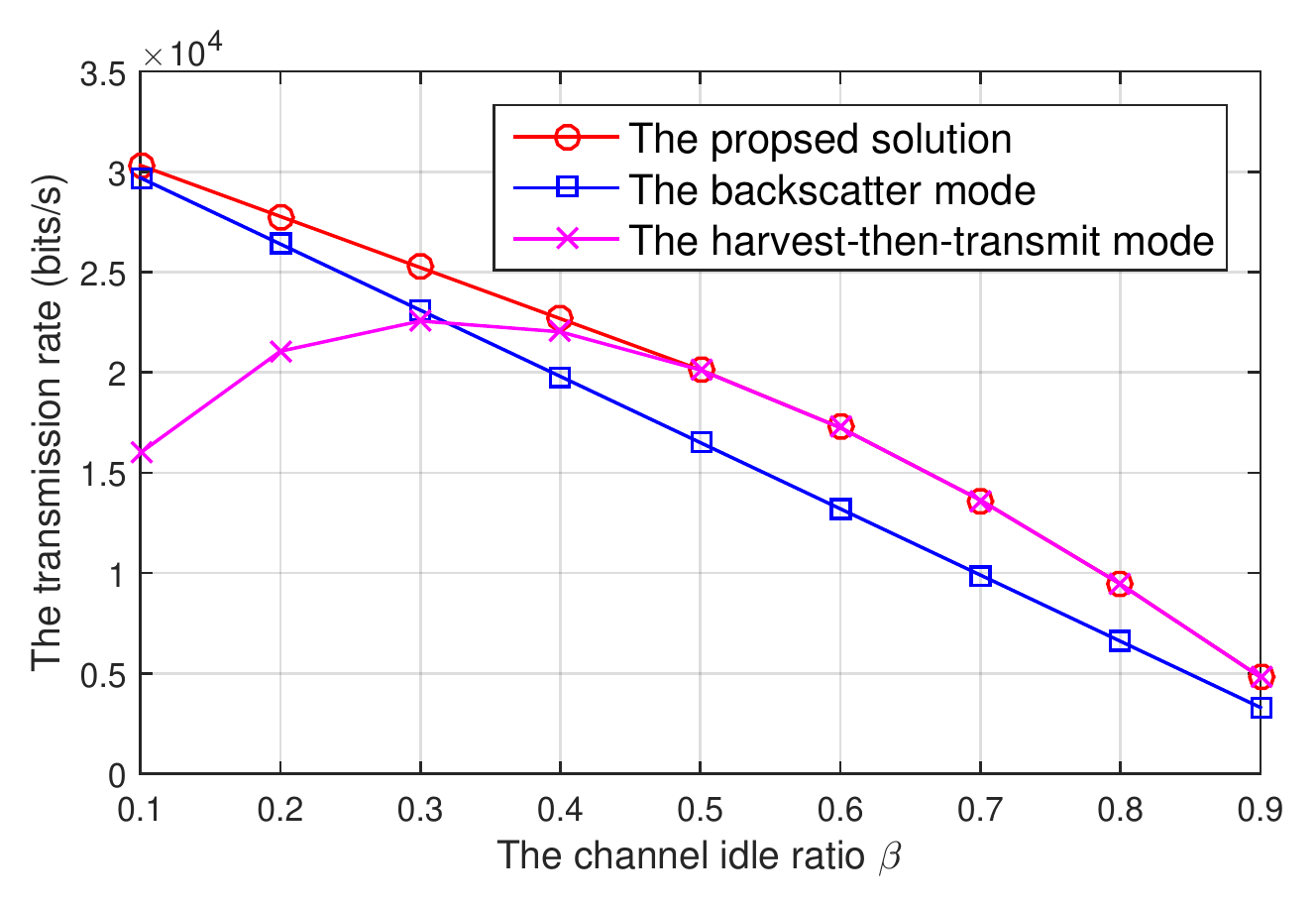}  \\
(b) \\ [-0.2cm] 
\end{array}$
\caption{The performance of the system under the variation of channel idle ratio.}
\label{fig:vary_beta}
\end{center}
\end{figure}

In Fig.~\ref{fig:Motivate} (on page 3), we show the variation of the objective function and the optimal value of $\alpha$. As shown in Fig.~\ref{fig:Motivate}, when $\alpha \in [\alpha^{\dagger},1]$, the objective function $R(\alpha)$ is concave, and it achieves the highest value at $\alpha^*=0.41125$. Then, in Figs.~\ref{fig:vary_beta}(a) and (b), we show the optimal value of $\alpha$ and the overall transmission rate of the secondary network when $\beta$ is varied. As shown in Fig.~\ref{fig:vary_beta}(a), as the channel idle ratio increases from $0.1$ to $0.6$, the optimal value of $\alpha$ gradually increases from $0.05$ to $1$. It remains at $1$ when the channel idle ratio is higher than $0.6$. This means that as the channel idle ratio increases, the ST will spend more time for harvesting energy instead of performing backscatter communication. Then, when the channel idle ratio is greater than or equal to $0.6$, the ST will always use the harvest-then-transmit mode. The reason is that the harvest-then-transmit mode can provide higher transmission rate than that of the backscatter mode. Thus, when the channel idle ratio is high, i.e., the busy channel period is small, the ST will spend the whole time to harvest energy when the channel is busy. Consequently, more bits can be transmitted during the channel idle period. 

In Fig.~\ref{fig:vary_beta}(b), we show the overall transmission rate obtained by the proposed solution. We compare the optimal results with two baseline strategies, i.e., backscatter mode (BM) and harvest-then-transmit mode (HM). Note that in BM or HM, the ST will only perform backscatter communication or energy harvesting, respectively. As shown in Fig.~\ref{fig:vary_beta}(b), the proposed solution always achieves the highest transmission rate compared with the BM and HM. In particular, when the channel idle ratio is $0.1$, the overall transmission rate obtained by the proposed solution is approximately $2$ times greater than those of the HM. When the channel idle ratio is $0.6$, the overall transmission rate obtained by the proposed solution is equal to that of the HM and almost $1.3$ times greater than that of the BM. Note that when the channel idle ratio increases from $0.4$ to $0.9$, there is a decrease of transmission rate obtained by the HM. The reason is that when the channel idle ratio is too high, there is no energy for the ST to harvest, and thus the transmission rate will be reduced. 

\begin{figure}[!]
\begin{center}
$\begin{array}{c} 
\epsfxsize=2.6 in \epsffile{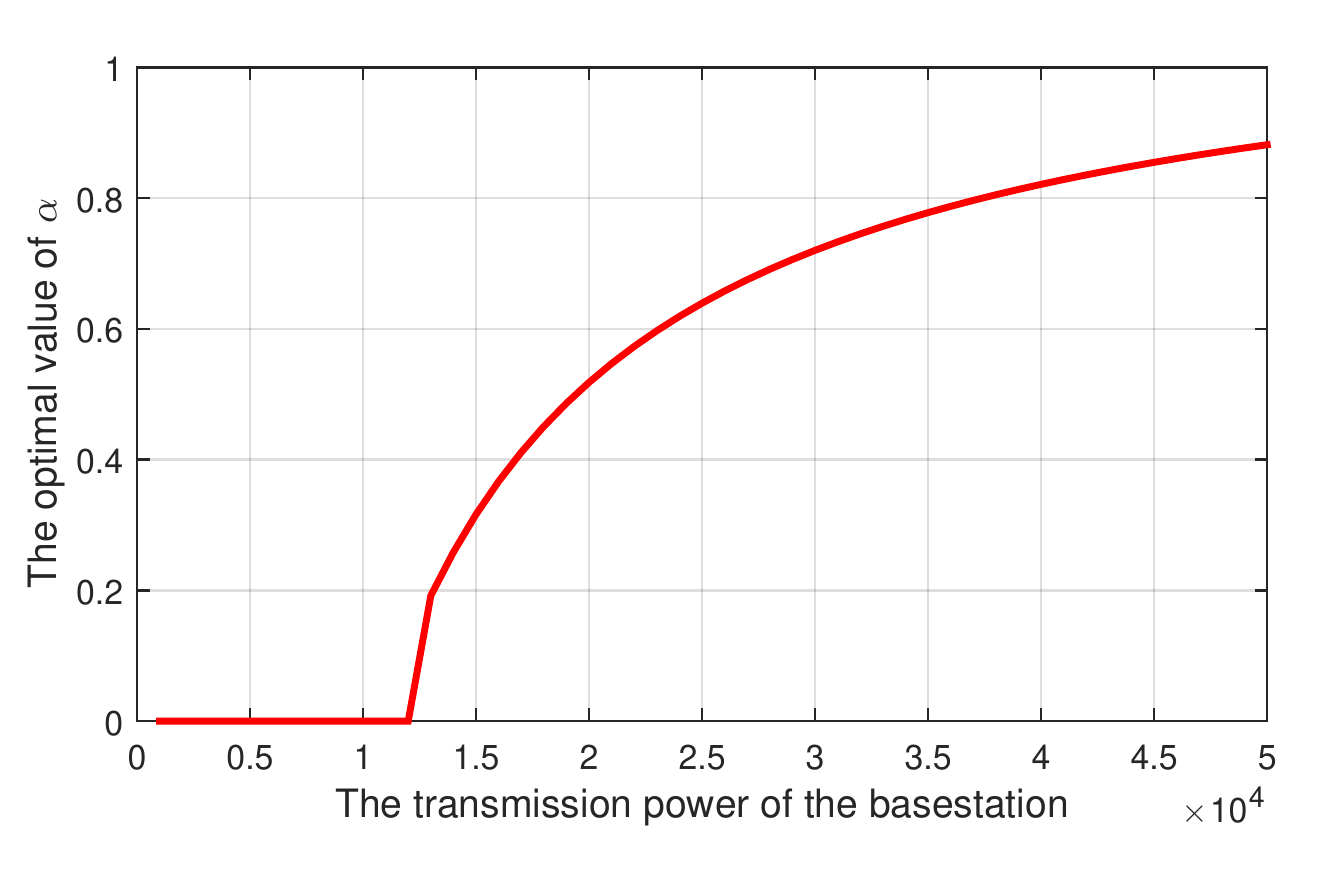}   \\
(a) \\
\epsfxsize=2.6 in \epsffile{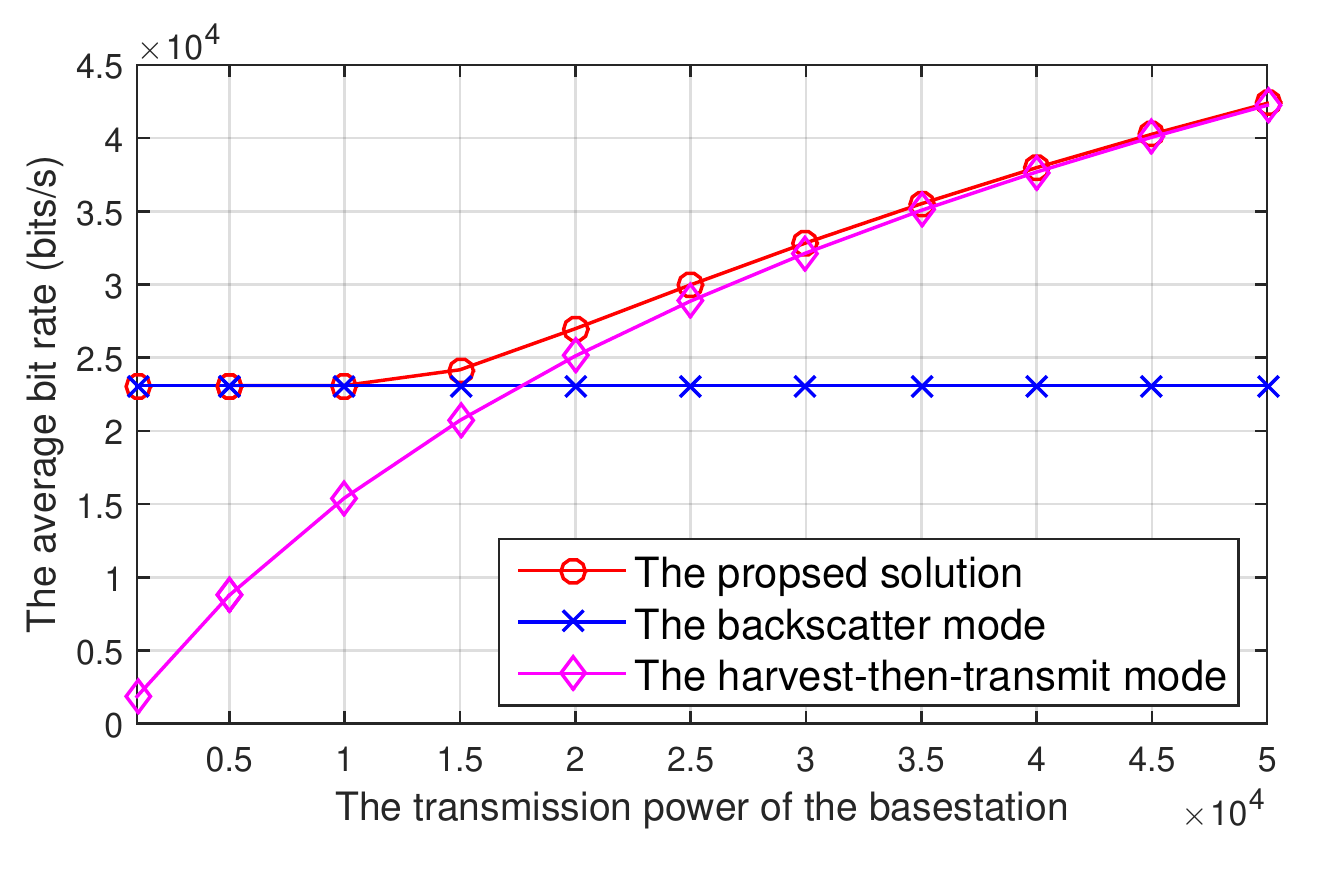}  \\
(b) \\ [-0.2cm]
\end{array}$
\caption{The performance of the system under the variation of transmission power of the base station.}
\label{fig:vary_PT}
\end{center}
\end{figure}

\begin{figure}[!]
\begin{center}
$\begin{array}{cc} 
\epsfxsize=2.7 in \epsffile{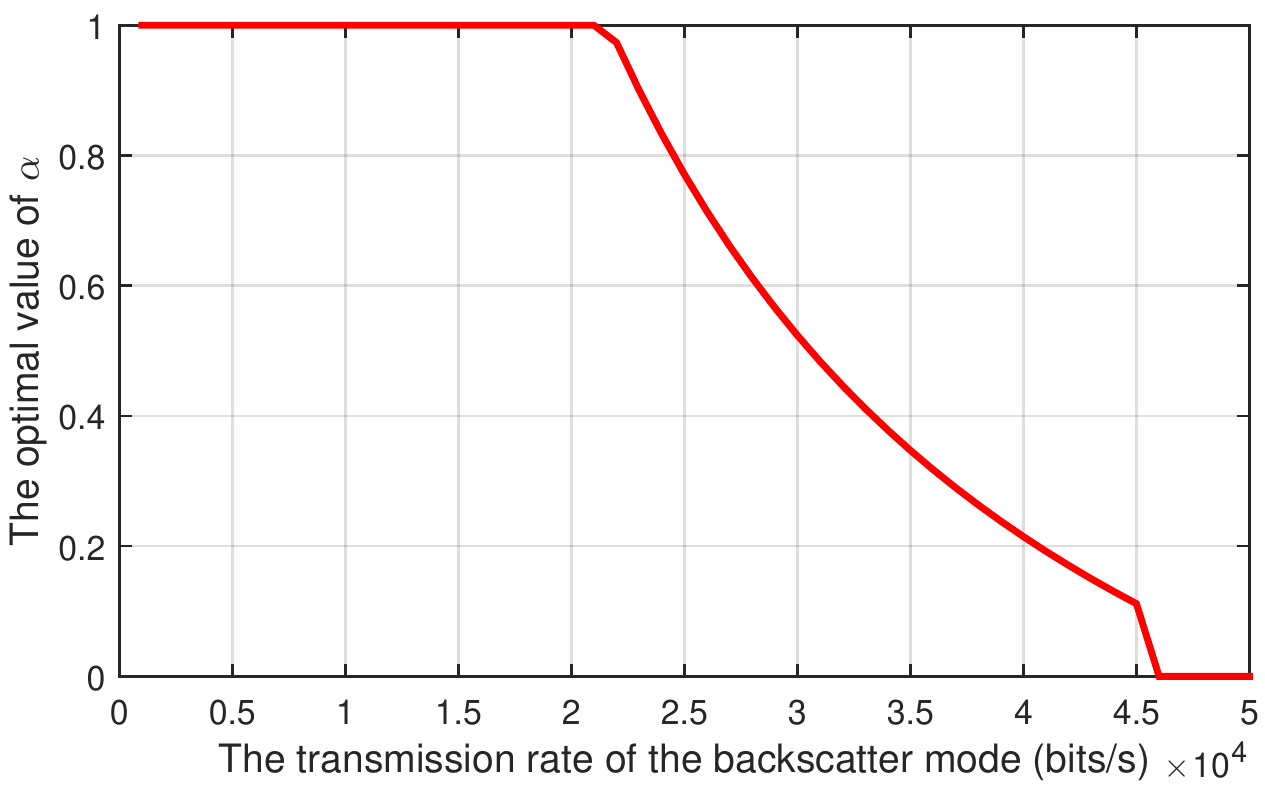}   \\
(a) \\
\epsfxsize=2.7 in \epsffile{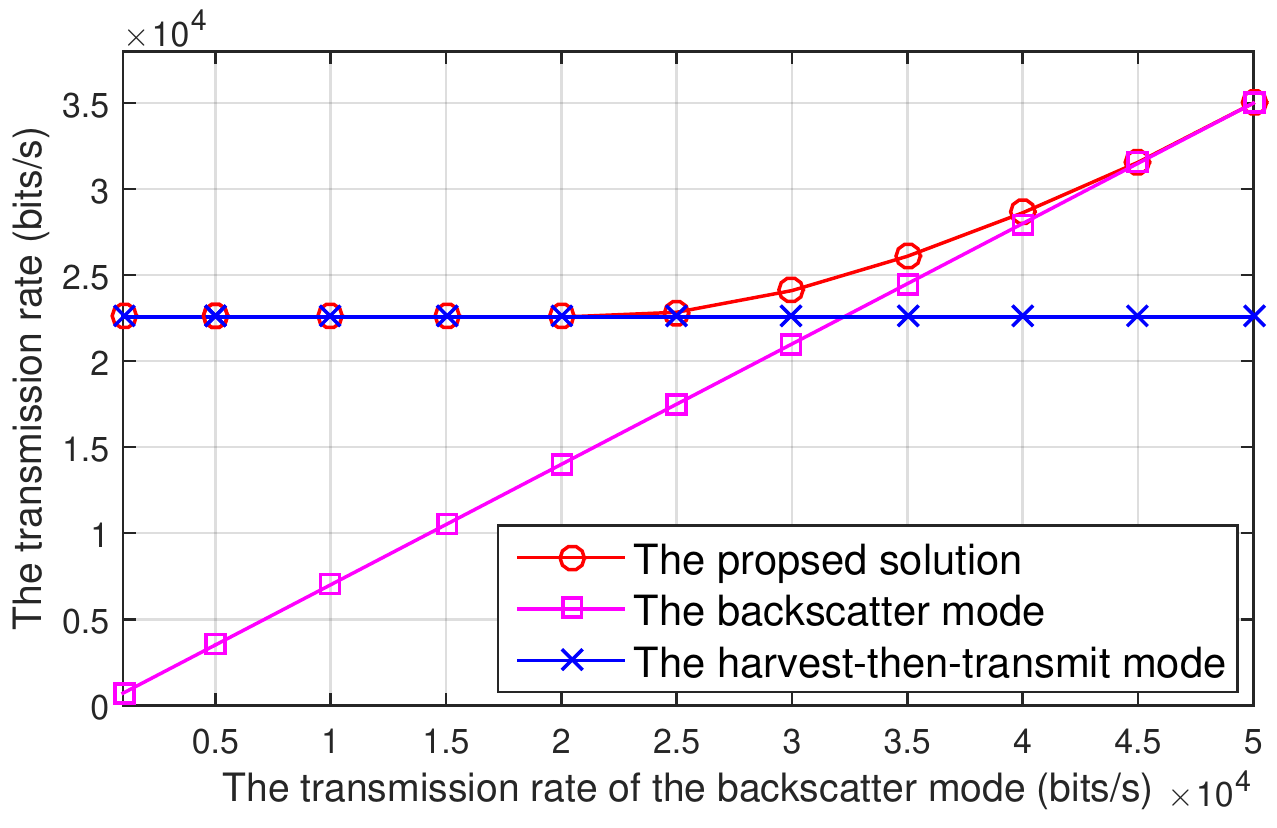}  \\
(b) \\ [-0.2cm]
\end{array}$
\caption{The performance of the system under the variation of transmission rate of the backscatter mode.}
\label{fig:vary_Bb}
\end{center}
\end{figure}

We then vary the transmission power of the PT (Fig.~\ref{fig:vary_PT}) and the transmission rate of the BM (Fig.~\ref{fig:vary_Bb}) to evaluate the performance as well as the tradeoff between BM and HM modes. In Fig.~\ref{fig:vary_PT} (a), as the transmission power of the PT increases, the optimal value of $\alpha$ remains at zero (i.e., the BM) when the transmission power is lower than $13$kW, and it increases gradually to $0.9$ as the transmission power increases from $13$kW to $50$kW. In Fig.~\ref{fig:vary_Bb}(a), as the transmission rate of the backscatter mode increases, the optimal value of $\alpha$ remains at one (i.e., the HM) when the transmission rate of the BM is lower than $21$kbps, and it will be reduced gradually, i.e., the ST tends to spend more time for the backscatter mode, to zero when the transmission rate of the BM is greater than $45$kbps. Furthermore, as shown in both Fig.~\ref{fig:vary_PT}(b) and Fig.~\ref{fig:vary_Bb}(b), the overall transmission rate obtained by the proposed solution always achieves the best performance compared with that of the BM and the HM. 

\section{Summary} 
In this work, we have proposed a new concept of integrating the ambient backscatter communication with RF-powered CRN. We then have introduced an optimization problem to obtain an optimal solution for the secondary transmitter to backscatter signals or to harvest energy for data transmission. The objective is to maximize the overall transmission rate for the secondary network. Numerical results have shown that our proposed solution can achieve significantly better performance compared with using either backscatter communication or harvest-then-transmit protocol. 

\appendices

\section{The proof of Theorem~\ref{theo:opt_data_trans_period}}
\label{app:theo:opt_data_trans_period}

Since $\alpha^{\dagger} \leq 1$ and $\alpha^{\dagger} \leq \alpha$, then from~(\ref{eq:R_h}), we have
\begin{equation}
	R_{\mathrm{h}} = \mu \kappa W \log_2 \left[ 1 + \frac{1}{P_0 \mu} \big( \alpha(1-\beta)P_{\mathrm{R}} - E_{\mathrm{c}}	\big)	\right]. 
\end{equation}
To prove Theorem~\ref{theo:opt_data_trans_period}, we denote 
\begin{eqnarray}
a & = & \kappa W 	,	\\
b & = & \frac{1}{P_0} \Big( \alpha(1-\beta)P_{\mathrm{R}} - E_{\mathrm{c}}	\Big)	,
\end{eqnarray}
where $a$ and $b$ are positive constants since now we consider $R_{\mathrm{h}}$ as a function of $\mu$. Then, (\ref{eq:R_h}) becomes
\begin{equation}
R_{\mathrm{h}}(\mu) = a \mu \log_2 \left( 1 +  \frac{b}{\mu} \right)	.
\end{equation}
We then derive the first and second derivatives of $R_{\mathrm{h}}$ with respect to $\mu$ as follows:
\begin{eqnarray} 
R^{'}_{\mathrm{h}}(\mu) &=& a \log_2 \Big(1+\frac{b}{\mu} \Big)  -  \frac{ab}{(\mu+b)\ln 2}  	,	   \label{eq:first_differentiation_Rh} \\
R^{''}_{\mathrm{h}}(\mu) &=& -\frac{ab^2}{\mu(\mu+b)^2 \ln 2}  \label{eq:second_differentiation_Rh}	.
\end{eqnarray}

From~(\ref{eq:second_differentiation_Rh}), we show that $R^{''}_{\mathrm{h}} < 0$ since $a$, $b$, and $\mu$ are greater than 0. Hence, $R^{'}_{\mathrm{h}}(\mu)$ is a decreasing function with respect to $\mu$. Moreover, from (\ref{eq:first_differentiation_Rh}), we derive the following result
\begin{equation}
\begin{aligned}
\lim_{\mu \rightarrow +\infty} R^{'}_{\mathrm{h}}(\mu) & = \lim_{\mu \rightarrow +\infty} a \log_2 \Big(1+\frac{b}{\mu} \Big) 	- 	\lim_{\mu \rightarrow +\infty}	 \frac{ab}{(\mu+b)\ln 2} 	,	\\
& = 	0.
\end{aligned}
\end{equation}
This implies that $R^{'}_{\mathrm{h}}(\mu) >0, \forall \mu \in [0,\beta]$. As a result, $R_{\mathrm{h}}(\mu)$ is an increasing function over $\mu \in [0,\beta]$, and thus $\max_{\mu} R_{\mathrm{h}} (\mu) = R_{\mathrm{h}}(\beta), \forall \mu \in [0,\beta]$.

The proof now is completed.

\section*{Acknowledgment}
This work was supported in part by the National Research Foundation of Korea (NRF) grant funded by the Korean government (MSIP) (2014R1A5A1011478), Singapore MOE Tier 1 (RG18/13 and RG33/12) and MOE Tier 2 (MOE2014-T2-2-015 ARC4/15 and MOE2013-T2-2-070 ARC16/14), and the U.S. National Science Foundation under Grants US NSF ECCS-1547201, CCF-1456921, CNS-1443917, ECCS-1405121, and NSFC61428101.

\bibliographystyle{IEEE}

\end{document}